\documentclass[conference]{IEEEtran}
\usepackage{cite}

\ifCLASSINFOpdf
  \usepackage[pdftex]{graphicx}
\else
\fi
\usepackage[cmex10]{amsmath}
\usepackage{url}

\newcommand{\tofab}{T_{a\leftrightarrow b}}
\newcommand{\tofat}{T_{a\leftrightarrow l}}
\newcommand{\tofbt}{T_{b\leftrightarrow l}}

\newcommand{\hattofab}{\hat{T}_{a\leftrightarrow b}}

\newcommand{\tdoa}{TD}
\newcommand{\hattdoa}{\hat{TD}}

\begin{document}
%

\title{Time Difference of Arrival Extraction\\ from Two-Way Ranging}


\author{\IEEEauthorblockN{Patrick Rathje\textsuperscript{1}, Olaf Landsiedel\textsuperscript{1,2}} 
\IEEEauthorblockA{
\textsuperscript{1}\textit{Kiel University, Germany} \\
\textsuperscript{2}\textit{Chalmers University of Technology, Sweden}\\
\{pra, ol\}@informatik.uni-kiel.de}
}


%


\maketitle

\begin{abstract}
Two-Way Ranging enables the distance estimation between two active parties and allows time of flight measurements despite relative clock offset and drift. Limited by the number of messages, scalable solutions build on Time Difference of Arrival (TDoA) to infer timing information at passive listeners. However, the demand for accurate distance estimates dictates a tight bound on the time synchronization, thus limiting scalability to the localization of passive tags relative to static, synchronized anchors. This work generalizes the extraction of Time Difference of Arrival information from a Two-Way Ranging process and enables passive tags to extract distance information from a ranging process, allowing scalable tag localization without the need for static or synchronized anchors.
The error due to clock drifts is formally deducted and formulas for the correction of relative clock drifts are derived. The introduced correction formulas reduce the estimation error to the clock drift of one party, resulting in accurate TDoA measurements despite relative clock offset and drift for the Double-Sided Two-Way Ranging and with additional carrier frequency offset estimation, also for Single-Sided Two-Way Ranging.
\end{abstract}



%
\IEEEpeerreviewmaketitle

\section{Introduction}
Measuring the time of flight (ToF) of wireless packets allows devices to estimate their respective distances.
The rise of cheap and small ultra-wideband (UWB) transceivers brings this timing-based approach to small and mobile devices, enabling, e.g., indoor real-time location systems in which mobile tags are localized relative to static anchors.
Limited by the amount of channel utilization, dense deployments can build upon time differences on arrival (TDoA), allowing the localization of an unlimited number of tags but requiring sufficiently synchronized clocks of static anchors. Hence, scalable solutions like Talla or SnapLoc employ static anchors with a wired backbone for optimal performance~\cite{8911790, grossiwindhager2019snaploc}.

For range estimation in dynamic scenarios, the Two-Way Ranging (TWR) approach allows two active nodes to estimate their distance irrespective of clock offsets by comparing relative time intervals. However, even minor clock deviations challenge the resulting estimation accuracy, with each nanosecond of clock deviation resulting in approximately 30~cm error. In TWR, the relative drift between the active devices remains a major source of error. In case of the traditional Single-Sided Two-Way Ranging (SS-TWR), the carrier frequency offset (CFO) estimation allows a correction of the relative clock drift~\cite{8555809}. Instead of CFO, the Double-Sided Two-Way Ranging (DS-TWR) adds another message to correct relative drifting. With the assumption of a fixed clock drift during the execution of the protocol, both variants can theoretically limit the influence to the drift of one participant.

The TWR and TDoA approaches can be combined to Passive Extended DS-TWR~\cite{7993831} respectively Active-Passive TWR~\cite{9309999}. In these systems, a tag performs TWR with active anchors while additional, passive anchors listen to the exchange. Based on the extracted TDoA values and known distances between anchors, the system then infers the distance between the tag and passive anchors. However, only the CFO estimation reduced the systematic error down to the clock drift of a single device~\cite{8555809}. Further on, tags remain active, thus limiting scalability.

\begin{figure}
    \centering
    \includegraphics[width=0.8\linewidth]{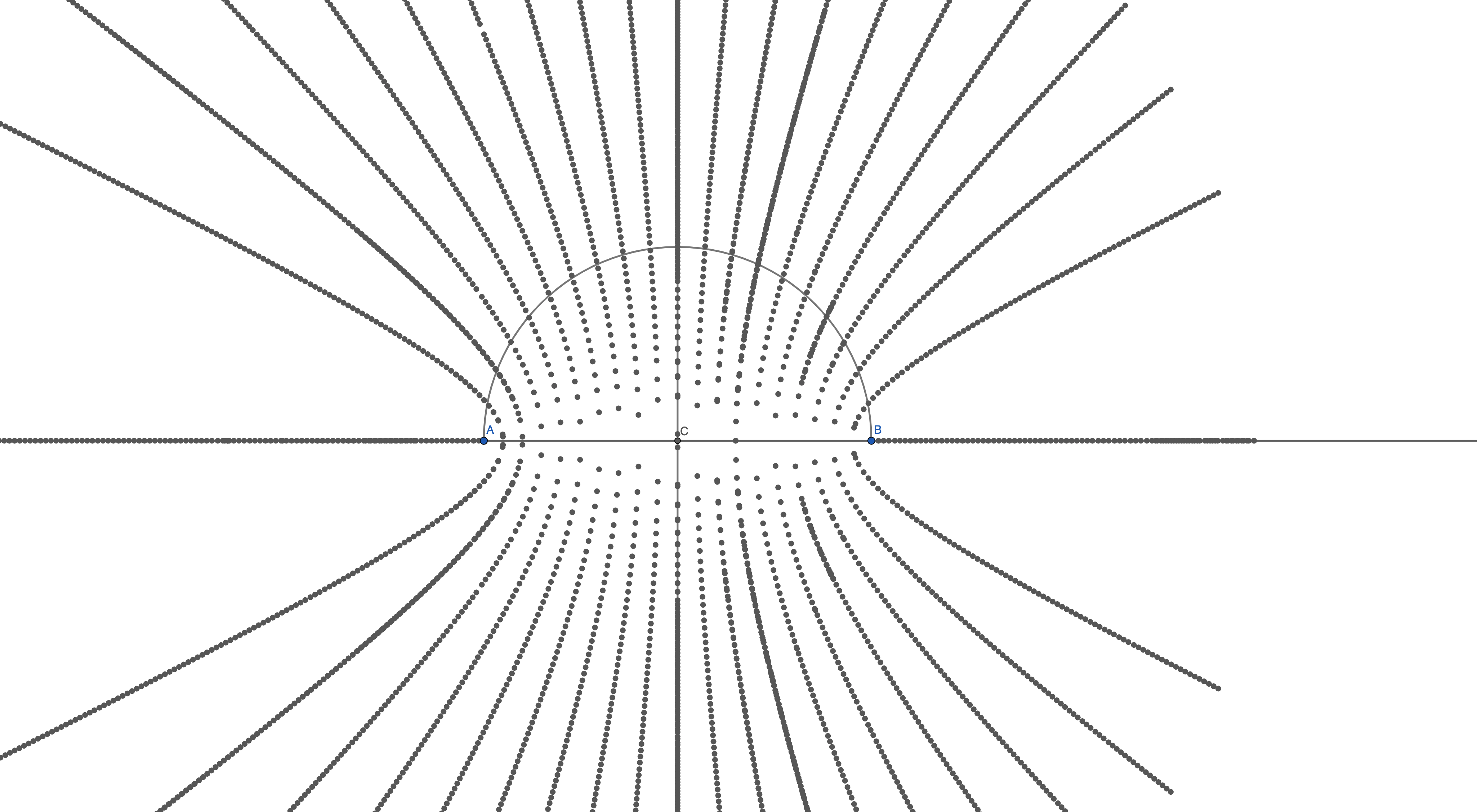}
    \caption{
    TDoA inferred Hyperbolas: The time difference of arrival (TDoA) allows a tag to fix its position to a hyperbola relative to known anchors, allowing hyperbolic positioning with enough anchors. This work describes how TDoA information can be extracted from listening to the Two-Way Ranging of other devices, allowing for mobile anchors and an increase in the frequency and scalability for dense deployments.
    }
    \label{fig:offset-information}
\end{figure}

We find that an active TWR exchange allows all listeners to extract the timing difference relative to the active parties, i.e. without the need for known distances. Hence, TDoA extraction can be generalized to passive tags as well. As Figure \ref{fig:offset-information} illustrates, this generalization allows a passive tag to limit its position estimate to a hyperbola relative to the active participants, enabling hyperbolic positioning. In addition, errors due to relative clock drifts can be mitigated even on passive devices. This work therefore revisits and generalizes TDoA extraction from Two-Way Ranging. Covering both the Single- and Double-Sided TWR variant, systematic errors due to relative clock drifts are analyzed and corrected. Overall, this work contributes the following:
\begin{enumerate}
    \item We introduce a common error estimation and correction formula for SS- and DS-TWR under relative clock drifts.
    \item We describe the generalized extraction of TDoA information from listening to a TWR ranging process.
    \item We analyze as well as correct the error under relative clock drifts for both SS-TWR and DS-TWR (with and without CFO estimation respectively).
\end{enumerate}

The structure of this work is as follows: Section \ref{sec:twr} introduces the measurement model, the mathematical notation and our unified error analysis and correction for the traditional TWR under relative clock drifts. Then, Section \ref{sec:tdoa} introduces the passive TDoA extraction with the corresponding error estimation and correction. Section \ref{sec:related-work} summarizes related work. Finally, Section \ref{sec:conclusion} concludes this work.

\section{Two-Way Ranging}
\label{sec:twr}
\begin{figure}
    \centering
    \includegraphics[width=0.8\linewidth, clip, trim=0.0cm 0cm 1.5cm 0cm]{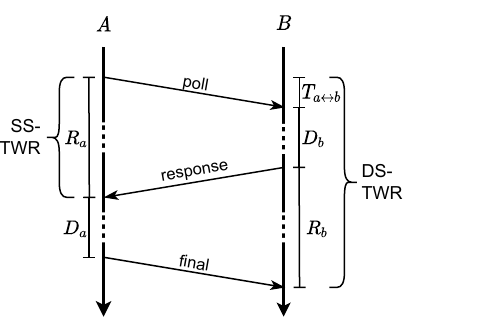}
    \caption{The message exchange in Two-Way Ranging (TWR) protocols: Measuring the round times and delays, active parties $A$ and $B$ can estimate the time-of-flight and therefore their distance despite clock offsets. The addition of a third (final) message extends the traditional Single-Sided TWR (SS-TWR) to Double-Sided TWR (DS-TWR) and mitigates the effect of relative clock drifts removing the need for carrier offset estimation as in SS-TWR.
    }
    \label{fig:twr}
\end{figure}

The traditional or Single-Sided Two-Way Ranging (SS-TWR) computes the distance based on one poll and one response message in each direction, measuring the overall round time for the exchange as well as the delay of the other party. Double-Sided Two-Way Ranging (DS-TWR) bases on two interleaved Single-Sided Two-Way Ranging (SS-TWR) rounds:
Each side initializes one round with a message while the other responds. As one message is shared between both rounds, three messages are exchanged in total. Figure \ref{fig:twr} displays the message exchange for SS-TWR and DS-TWR.

For the ranging of two participants $A$ and $B$ we denote the respective round time as $R_a$ and the response delays as $D_b$. Let $\tofab$ denote the time-of-flight between participants $A$ and $B$.  Without loss of generality let $A$ initiate the TWR process with $B$ and further assume that $A$ is computing the distance estimate.
Moreover, we assume that messages are timestamped before their transmission. Otherwise, a downstream message could deliver the missing timestamps.

\subsection{TWR without Clock Drift}
Assuming that clocks run with offsets but do not experience different drifts, after the ranging process, $A$ can determine the ToF $\tofab$, as it relates to the respective round times as follows:

\begin{equation}
R_a = 2 \tofab + D_b
\label{ra_to_tofab}
\end{equation}

As both $R_a$ and $D_b$ are durations, both TWR variants can successfully estimate the accurate distance despite clock offsets. Based on Equation \eqref{ra_to_tofab}, for the ToF between $A$ and $B$ it holds:

\begin{equation}
\tofab = \frac{R_a - D_b}{2}
\label{ss_tofab}
\end{equation}

However, due to clock imperfections, the clocks tick with different frequencies, probably drifting apart while executing the protocol. 

\subsection{Raw TWR with Clock Drift}
We denote the clock drift for $A$ as $e_a$ resulting in the factor $k_a$ with $k_a := 1+e_a$ and define $\hat{X}$ as the locally measured value of $X$~\cite{7822844}:

$$\hat{R}_a= (1+e_a) R_a = k_a R_a$$
$$\hat{D}_b= (1+e_b) D_b = k_b D_b$$

For the DS-TWR variant we further define:
$$\hat{D}_a= (1+e_a) D_a = k_a D_a$$
$$\hat{R}_b= (1+e_b) R_b = k_b R_b$$

We also denote the estimate of $\tofab$ as $\hattofab$.
Plugging our locally measured values directly into Equation \eqref{ss_tofab} yields the raw ToF estimate for TWR:

\begin{equation}
\hattofab^{\text{RAW}} := \frac{\hat{R}_a - \hat{D}_b}{2} \label{ss_cfo}
\end{equation}

However, without further correction the clock drifts induce the following error:

\begin{align}
\hattofab^{\text{RAW}} - \tofab &= \frac{\hat{R}_a - \hat{D}_b}{2} - \frac{R_a - D_b}{2} \nonumber\\
  &= \frac{\hat{R}_a - R_a  + D_b - \hat{D}_b  }{2} \nonumber\\
  &= \frac{e_a R_a  - e_b D_b}{2} \label{raw_twr_error}
\end{align}

Our raw error thus depends on the drifts influencing the components $R_a$ and $D_b$ differently. Note that even in the improbable case of $e_a = e_b$ we still get an error in our estimation.

\subsection{Drift-Corrected TWR}
In fact, with Equation \eqref{ra_to_tofab}, we can remodel the error of Equation \eqref{raw_twr_error} as follows:

\begin{align}
\hattofab^{\text{RAW}} - \tofab &= \frac{e_a R_a  - e_b D_b}{2} \nonumber\\
    &=\frac{e_a (2\tofab+D_b)  - e_b D_b}{2} \nonumber\\
    &= \frac{e_a 2 \tofab + e_a D_b  - e_b D_b}{2} \nonumber\\
    &= e_a \tofab + \frac{ e_a - e_b}{2} D_b \nonumber\\
    &= e_a \tofab + \frac{ k_a - k_b}{2} D_b \nonumber\\
    &= e_a \tofab + \frac{ k_a - k_b}{2} \frac{1}{k_b} \hat{D}_b \nonumber\\
    &= e_a \tofab + \frac{1}{2} \frac{k_a}{k_b} \hat{D}_b - \frac{1}{2} \hat{D}_b \label{ss_twr_error_todo}
\end{align}

Hence, the error stems from the drift influence on the actual ToF by $A$ and the relative drift factor $\frac{k_a}{k_b}$ resulting from different clock drifts of $A$ and $B$.
We can therefore correct our raw ToF estimation with the relative drift factor relative to the clock of $A$:

\begin{align}
\hattofab^{\text{DC-A}} :&= \hattofab^{\text{RAW}} - (\frac{1}{2} \frac{k_a}{k_b} \hat{D}_b - \frac{1}{2} \hat{D}_b) \nonumber\\
&=\frac{\hat{R}_a - \hat{D}_b}{2} - \frac{1}{2} \frac{k_a}{k_b} \hat{D}_b + \frac{1}{2} \hat{D}_b \nonumber\\
&=\frac{\hat{R}_a - \frac{k_a}{k_b} \hat{D}_b}{2} \label{twr_corrected_a}
\end{align}

Comparing this to the actual ToF gives:

\begin{align}
\hattofab^{\text{DC-A}}  &= \frac{\hat{R}_a - \frac{k_a}{k_b} \hat{D}_b}{2} \nonumber\\
&= \frac{k_a R_a - k_a D_b}{2} \nonumber\\
&= k_a \frac{ R_a - D_b}{2} \nonumber\\
&= k_a \tofab \label{twr_corrected_a_twr_relation}
\end{align}

So, when we can determine the relative drift factor $\frac{k_a}{k_b}$, we can bring our error down to just the drift during the ToF:

\begin{equation}
\hattofab^{\text{DC-A}} - \tofab = k_a \tofab - \tofab = e_a \tofab \nonumber
\end{equation}

As reordering Equation \eqref{ra_to_tofab} yields $D_b = R_a - 2 \tofab$, we can analogously derive the drift corrected ToF for the clock of $B$:
\begin{align}
\hattofab^{\text{DC-B}} :&= \hattofab^{\text{RAW}} - (\frac{1}{2} \hat{R}_a -  \frac{1}{2}  \frac{k_b}{k_a} \hat{R}_a) \nonumber\\
&= \frac{\hat{R}_a - \hat{D}_b}{2} - \frac{1}{2} \hat{R}_a  +  \frac{1}{2}  \frac{k_b}{k_a} \hat{R}_a \nonumber\\
&= \frac{\frac{k_b}{k_a} \hat{R}_a - \hat{D}_b}{2} \label{twr_corrected_b}
\end{align}
 
Using for example, Carrier Frequency Offset estimation (CFO) ~\cite{8555809}, we can determine the relative clock drift factor $\frac{k_a}{k_b}$.
As an alternative, the symmetry within DS-TWR can be exploited, requiring an additional message but no CFO estimation. Since assuming that both DS-TWR rounds incorporate the same ToF, it holds that:
\begin{equation}
R_a + D_a = D_b + R_b
\label{ra_da_eq_db_rb}
\end{equation}

Thus, for the actually measured intervals, we derive:
\begin{equation}
\frac{1}{k_a} (\hat{R}_a + \hat{D}_a) = \frac{1}{k_b} (\hat{D}_b + \hat{R}_b)\nonumber
\end{equation}

Which can be rearranged to estimate the relative drift factor:

\begin{equation}
\frac{k_a}{k_b}  = \frac{R_a + D_a}{D_b + R_b}
\label{alt_twr}
\end{equation}

This correction within the DS-TWR is known as the Alternative DS-TWR approach~\cite{7822844}. Plugging Equation \eqref{alt_twr} into \eqref{twr_corrected_a} (or \eqref{twr_corrected_b} respectively) yields the Alternative DS-TWR estimation formula.

\section{TDoA Extraction for Two-Way Ranging}
\label{sec:tdoa}

Let $L$ now be a passive listener to the ranging process, trying to estimate the Time Difference of Arrival (TDoA) between $A$ and $B$. This TDoA describes the relative difference of the arrival times as if $A$ and $B$ sent their messages at the same time, i.e. the difference in ToF to $A$ and $B$ modeled by $\tdoa := \tofat-\tofbt$. As neither in SS-TWR nor in DS-TWR are $A$ and $B$ sending at the same time, this TDoA is also called asynchronous (A-TDoA)~\cite{8555809}. Using their ToF $\tofab$ and the delays of the TWR process, $L$ can relate the respective durations: Assuming that $L$ receives the respective messages, it determines $M_a$ as its own experienced time difference between the poll issued by $A$ and the response message sent by $B$.

\subsection{Basic TDoA Extraction}
As the poll message from $A$ travels simultaneously to $L$ and $B$, $M_a$ starts $\tofat$ after the initial message from $A$ and ends $\tofbt$ after the response message from $B$. Including the delay $D_b$, for $M_a$, the following holds:

\begin{equation}
M_a = (\tofab-\tofat)  + D_b + \tofbt
\label{ss_ma_db_relation}
\end{equation}

$M_a$ thus depends on the ToF between the participants $A$ and $B$ but also their respective ToF to $L$. Rearranging Equation \eqref{ss_ma_db_relation} yields the wanted TDoA:

\begin{equation}
\tdoa = \tofab + D_b - M_a 
\label{td_by_db}
\end{equation}

Replacing $D_b$ with $R_a - 2 \tofab$ based on Equation \eqref{ra_to_tofab} gives:
\begin{align}
\tdoa &= \tofab + R_a - 2\tofab - M_a \nonumber \\
&= R_a - \tofab - M_a 
\label{td_by_ra}
\end{align}


\subsection{TDoA Relative Drift Correction}

Naturally, also the clock of $L$ drifts so that $L$ determines the local, potentially drifted time $\hat{M}_a$:

$$\hat{M}_a= (1+e_l) M_a = k_l M_a$$

As before, we can now determine the raw estimate:

\begin{align}
\hattdoa^{\text{RAW-A}} &:= \hat{R}_a - \hattofab - \hat{M}_a  \label{td_raw_a}
\end{align}

And its associated error:

\begin{align}
\hattdoa^{\text{RAW-A}} - \tdoa &= \hat{R}_a - \hattofab - \hat{M}_a  - (R_a - \tofab - M_a)  \nonumber \\
&= \hat{R}_a - \hattofab - \hat{M}_a - R_a + \tofab + M_a  \nonumber  \\
&= \hat{R}_a - R_a + \tofab - \hattofab  + M_a - \hat{M}_a \nonumber \\
&= (\tofab - \hattofab) + e_a R_a  - e_l M_a \label{td_raw_error_a}
\end{align}

Derived from \eqref{td_raw_error_a}, the error depends on the error in the estimated ToF as well as the relative clock drift between $A$ and $L$.
However, assuming knowledge of the relative drift factor $\frac{k_a}{k_l}$, we can relate $M_a$ to $R_a$ and correct the relative drift, i.e. based on the clock of $A$:

\begin{align}
\hattdoa^{\text{DC-A}} &:= \hat{R}_a - \hattofab^{\text{DC-A}} - \frac{k_a}{k_l} \hat{M}_a  \label{td_dc_a} 
\end{align}

Obviously, we can also estimate the difference from the perspective of the clock from $L$, also correcting the ToF estimate accordingly:
\begin{align}
    \hattdoa^{\text{DC-L}} &:= \frac{k_l}{k_a} (\hat{R}_a - \hattofab^{\text{DC-A}}) - \hat{M}_a \label{td_dc_t}
\end{align}

\begin{figure}
    \centering
    \includegraphics[width=0.8\linewidth, clip, trim=0.0cm 0cm 2.5cm 0cm]{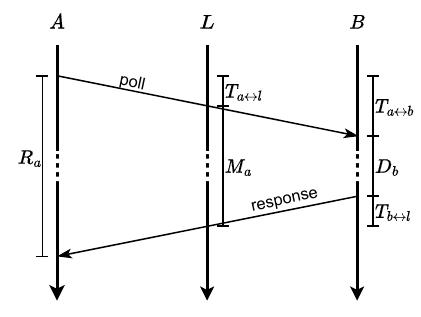}
    \caption{ Inferred TDoA from SS-TWR: A passive listening node $L$ determine the difference in ToF between nodes $A$ and $B$ which conducted active ranging using SS-TWR. Estimating the carrier frequency offset mitigates the effect of clock drifts. 
    }
    \label{fig:ss-twr}
\end{figure}

It is important to mention that also $L$ can calculate $\hattdoa^{\text{DC-A}}$ given the necessary measurements as well as $\frac{k_l}{k_a}$ and $\frac{k_a}{k_b}$. As with the SS-TWR, $L$ could estimate the CFO to approximate the relative drift factors (see Figure \ref{fig:ss-twr}).
Estimating $\frac{k_l}{k_a}$ and $\frac{k_l}{k_b}$ allows $L$ to further compute $\frac{k_a}{k_b}$ as follows:

\begin{equation}
   \frac{k_a}{k_b} = \frac{k_l}{k_b} \frac{k_a}{k_l}
\end{equation}

\begin{figure}
    \centering
    \includegraphics[width=0.8\linewidth, clip, trim=0.5cm 0cm 2.5cm 0cm]{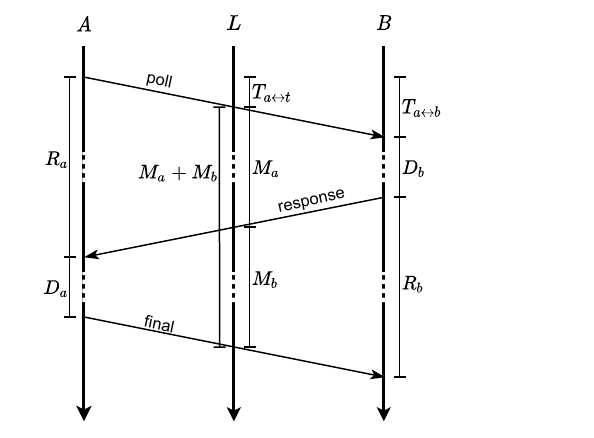}
    \caption{Inferred TDoA from DS-TWR: A passive listening node $L$ can determine the difference in time-of-flight nodes $A$ and $B$ which conduct active ranging using DS-TWR.  The two messages of $A$ allow $L$ to correct its relative clock drift.
    }
    \label{fig:ds-twr}
\end{figure}

In case that CFO estimation is not possible, the DS-TWR approach includes a second message from $A$, which finishes the second round. As Figure \ref{fig:ds-twr} displays, $L$ can thus measure the times $M_a$ and $M_b$ for both rounds respectively. Assuming that $\tofat$ stays equal during protocol execution, it holds that:

\begin{equation}
    R_a+D_a=M_a+M_b
\end{equation}

Similarly to Equation \eqref{ra_da_eq_db_rb}, $L$ can then derive the relative clock drift factor as follows:
\begin{equation}
\frac{k_l}{k_a} = \frac{\hat{M}_a+\hat{M}_b}{\hat{R}_a + \hat{D}_a}\nonumber
\end{equation}

Transposing Equation \eqref{td_dc_a} and using \eqref{twr_corrected_a_twr_relation} reveals the drift influence:

\begin{align}
    \hattdoa^{\text{DC-A}} &= \hat{R}_a - \hattofab^{\text{DC-A}} - \frac{k_a}{k_l} \hat{M}_a \nonumber \\
    &= \hat{R}_a - k_a \tofab - \frac{k_a}{k_l} \hat{M}_a \nonumber \\
    &= k_a R_a - k_a \tofab - \frac{k_a}{k_l} k_l M_a \nonumber \\
    &= k_a R_a - k_a \tofab - k_a M_a \nonumber \\
    &= k_a (R_a - \tofab - M_a) \nonumber \\
    &= k_a \tdoa  \\
\end{align}

Thus, the expected error in the drift-corrected case in times measured by $A$ is reduced as follows: 
\begin{align}
    \hattdoa^{\text{DC-A}} - \tdoa = k_a \tdoa - \tdoa =  e_a \tdoa \label{td_dc_a_error}
\end{align}

\section{Related Work}
\label{sec:related-work}
TDoA measurements constitute the foundation for scalable ranging systems. With its support for an unlimited number of passive devices, TDoA mechanisms are found in Global Navigation Satellite Systems as well as in UWB-based indoor localization~\cite{8911790, grossiwindhager2019snaploc, 9309999}.

With their Passive Extension to DS-TWR, Horváth, Ill and Milankovich devise a positioning algorithm that employs TDoA measurements from passive anchors~\cite{7993831}. They apply an extended variant of DS-TWR and reduce the influence of relative clock drifts (i.e., the systematic error). Following the multiplication idea of the alternative DS-TWR method~\cite{7822844}, Horváth et al. further reduce the systematic error for TDoA measurements~\cite{horvath2017passive} but do not decrease it to the drift of a single clock, as shown in this work.

Dotlic, Connell and McLaughlin use the carrier frequency offset estimation within Decawave’s DW1000 module to estimate and mitigate the relative clock offset. They further derive a CFO-based correction formula for TDoA extraction on passive anchors. In comparison, this work extends the TDoA correction to passive tags and further provides the correction for DS-TWR scenarios.

\section{Conclusion}
\label{sec:conclusion}
Using Time Difference of Arrival, listening nodes can increase their positional accuracy or even undertake hyperbolic positioning for fully passiveness, increasing scalability or the frequency of ranging systems. This work described and analyzed the extraction of TDoA information from the TWR process, enabling TDoA measurements without the need for static or synchronized anchors. Using the drift-corrected formulas of Equations \eqref{td_dc_a} and \eqref{td_dc_t}, the measurement error can be reduced to the effect of the clock drift from one participant and is thus negligible compared to the accuracy and range of technologies like UWB.

\bibliographystyle{IEEEtranS}
\bibliography{IEEEabrv,references}

\begin{thebibliography}{1}
\providecommand{\url}[1]{#1}
\csname url@samestyle\endcsname
\providecommand{\newblock}{\relax}
\providecommand{\bibinfo}[2]{#2}
\providecommand{\BIBentrySTDinterwordspacing}{\spaceskip=0pt\relax}
\providecommand{\BIBentryALTinterwordstretchfactor}{4}
\providecommand{\BIBentryALTinterwordspacing}{\spaceskip=\fontdimen2\font plus
\BIBentryALTinterwordstretchfactor\fontdimen3\font minus
  \fontdimen4\font\relax}
\providecommand{\BIBforeignlanguage}[2]{{%
\expandafter\ifx\csname l@#1\endcsname\relax
\typeout{** WARNING: IEEEtranS.bst: No hyphenation pattern has been}%
\typeout{** loaded for the language `#1'. Using the pattern for}%
\typeout{** the default language instead.}%
\else
\language=\csname l@#1\endcsname
\fi
#2}}
\providecommand{\BIBdecl}{\relax}
\BIBdecl

\bibitem{8555809}
I.~Dotlic, A.~Connell, and M.~McLaughlin, ``Ranging methods utilizing carrier
  frequency offset estimation,'' in \emph{2018 15th Workshop on Positioning,
  Navigation and Communications (WPNC)}, 2018, pp. 1--6.

\bibitem{grossiwindhager2019snaploc}
B.~Gro{\ss}iwindhager, M.~Stocker, M.~Rath, C.~A. Boano, and K.~R{\"o}mer,
  ``Snaploc: An ultra-fast uwb-based indoor localization system for an
  unlimited number of tags,'' in \emph{2019 18th ACM/IEEE International
  Conference on Information Processing in Sensor Networks (IPSN)}.\hskip 1em
  plus 0.5em minus 0.4em\relax IEEE, 2019, pp. 61--72.

\bibitem{7993831}
K.~A. Horv{\'a}th, G.~Ill, and {\'A}.~Mil{\'a}nkovich, ``Passive extended
  double-sided two-way ranging algorithm for uwb positioning,'' in \emph{2017
  Ninth International Conference on Ubiquitous and Future Networks
  (ICUFN)}.\hskip 1em plus 0.5em minus 0.4em\relax IEEE, 2017, pp. 482--487.

\bibitem{horvath2017passive}
------, ``Passive extended double-sided two-way ranging with alternative
  calculation,'' in \emph{2017 IEEE 17th International Conference on Ubiquitous
  Wireless Broadband (ICUWB)}.\hskip 1em plus 0.5em minus 0.4em\relax IEEE,
  2017, pp. 1--5.

\bibitem{9309999}
T.~Laadung, S.~Ulp, M.~M. Alam, and Y.~Le~Moullec, ``Active-passive two-way
  ranging using uwb,'' in \emph{2020 14th International Conference on Signal
  Processing and Communication Systems (ICSPCS)}, 2020, pp. 1--5.

\bibitem{7822844}
D.~Neirynck, E.~Luk, and M.~McLaughlin, ``An alternative double-sided two-way
  ranging method,'' in \emph{2016 13th Workshop on Positioning, Navigation and
  Communications (WPNC)}, 2016, pp. 1--4.

\bibitem{8911790}
D.~Vecchia, P.~Corbalán, T.~Istomin, and G.~P. Picco, ``Talla: Large-scale
  tdoa localization with ultra-wideband radios,'' in \emph{2019 International
  Conference on Indoor Positioning and Indoor Navigation (IPIN)}, 2019, pp.
  1--8.

\end{thebibliography}
%


\end{document}